\documentclass[%
 reprint,
 amsmath,amssymb,
 aps,
]{revtex4-1}

\usepackage{graphicx}
\usepackage{dcolumn}
\usepackage{bm}

\begin{document}

\preprint{APS/123-QED}

\title{Cavity and background oscillations in Intrinsic Josephson Junctions}%

\author{I. G. Hristov, R. D. Hristova and S. N. Dimova}
\affiliation {Faculty of Mathematics and Informatics, St. Kliment Ohridski University of Sofia,
  5 James Bourchier Blvd., 1164 Sofia, Bulgaria,
  \\emails: ivanh@fmi.uni-sofia.bg, radoslava@fmi.uni-sofia.bg, dimova@fmi.uni-sofia.bg }

\date{\today}

\begin{abstract}
Starting with zero initial conditions we simulate the
current-voltage characteristics (CVCs) of Intrinsic Josephson Junctions (IJJ)
for parameters close to that
of the BSCCO mesas. The simulation shows  a regular
pattern with voltage jumps at voltages satisfying the cavity resonance
conditions. The calculated emission has peaks at every voltage
jump. Analyzing  the  phases we observe a two-dimensional standing wave
pattern with a kink configuration as their static part.  Depending on the
kink configuration and the bias current, the kink amplitudes may differ from $\pi$ .
By means of Fast Fourier Transform (FFT) we show that the frequencies excellently satisfy  the ac
Josephson relation  as in the real experiments. Integer higher harmonics
up to  fifth order are also observed. Calculated amplitude maps
for the frequencies achieved and their harmonics demonstrate that together with cavity modes
there  exist nonlinear background modes. Direct
observation of the phase dynamics shows a relation between  oscillations
of the background modes and the static kink configuration: opposite static
kinks in different junctions correspond to opposite ($\pi$-shifted)
oscillations of the background modes.

\begin{description}

\item[PACS numbers]

74.50.+r, 74.72.-h, 85.25.Cp, 07.05.Tp

\end{description}
\end{abstract}

\maketitle


\section{Introduction}

The powerful $THz$ radiation from
BSCCO mesas at zero magnetic field being reported in 2007 \cite{p1} starts a new stage
of intensive investigation of the Intrinsic Josephson Junctions (IJJ).
A big progress  has been  made in mesa preparation technology, as a consequence self-heating
effects and hot spot formations are significantly reduced \cite{p2}-\cite{p6}.
The last progress gives possibility to increase the number of junctions in mesas from
$\approx 500$ \cite{p1}  to  $\approx 3000$ \cite{p4}, the
radiation power from  $ \approx 0.5 \mu w$  \cite{p1} to  $ \approx 30 \mu w$  \cite{p2} and
the frequency  from  $\approx 0.85 THz$  \cite{p1} up to $ \approx 2.4 THz$ \cite{p6}.

Still, the mechanism of powerful radiation has not been fully  explained.
It has been generally accepted that
high radiation power could be achieved when the frequency $\omega$
satisfies the resonance condition:
 \begin{equation}\label{e1}
 \omega=\omega_{J}=\omega_{cav}.
 \end{equation}
This means that $\omega$ satisfies the ac Josephson relation  $\omega=\omega_{J}=2ev/\hbar$
and it simultaneously coincides with some cavity mode frequency: $\omega=\omega_{cav}$.
Here $e$ is the electric charge, $\hbar$ is Planck's constant and $v$ is the applied dc voltage.
Cavity mode frequency $\omega_{cav}$ depends on mesa geometry and dimensions \cite{p7}.
The  resonance condition \eqref{e1} is rather restrictive
and usually it is not exactly observed in the real experiments.
Although the ac Josephson relation is always excellently fulfilled in the experiments, it is not yet explained
why the frequency $\omega$ may differ from the corresponding cavity mode frequency $\omega_{cav}$ \cite{p4},\cite{p5}.
The observed ''dual'' source mechanism  in experimental works  \cite{p8}, \cite{p9} is
confirmed numerically in \cite{p10} but it's still not well understood.

There are two main concepts for the powerful radiation mechanism - the coupling to the resonance modes
is obtained by $\pi$ kinks  \cite{p11},\cite{p12} or by breather type self oscillation
(not necessarily with $\pi$ amplitude)  \cite{p13}.
It is important to mention that there is a  difference in the mathematical models used in  \cite{p11} and  \cite{p13}.
 Periodic boundary conditions in z-direction (stacking direction) are posed in \cite{p11} while the  boundary conditions
 in \cite{p13} are nonperiodic.
Although the parameters are almost the same, there is a  difference in the characteristic mode velocities \cite{p14}.
Particularly  the maximum (in-phase) plasma velocities significantly differ for usual simulations with about 10-20 junctions.
On the other hand the use of nonperiodic boundary conditions in \cite{p13}
makes the junctions  to be not identical, the amplitudes of oscillations in the external junctions lower then these in the internal ones and hence  the picture becomes more complicated.
Actually, it is reasonable to think that the solutions in \cite{p11} and \cite{p13} are generally of the same type and not to treat them as different.

In addition we want to note the new state with coexisting moving fluxons and longitudinal plasma wave in \cite{p15}.
This state may be also important for the future understanding of powerful THz radiation.

In this work we carry out numerical simulation by solving a system of 1D Sine-Gordon  equations which  describe well
the rectangular mesas in which the zero mode is realized in y-direction.
Because in reality the number of junctions is thousands, we  prefer to use periodic boundary conditions as in \cite{p11}.
We  achieve kinks  in the numerical simulations and perform  a detailed  analysis of their  possible amplitudes.
We show that the kink amplitudes may differ from $\pi$, they depend on the kink configuration in z-direction
and on the bias current. A crucial observation is the alternative character of the static kink configuration
and approximately the same number of kinks and anti-kinks in the stack.
We show by FFT  analysis that radiation frequencies excellently satisfy the ac Josephson relation as in the real
experiments. Calculated amplitude maps and the direct phase analysis show simultaneous existence of nonlinear
background modes. Although the background modes are  hard to see  at the top of the current steps,
 corresponding to a particular  cavity mode excitation  \cite{p11},
their amplitudes become higher when the current decreases. The background modes  are very well seen at the bottom
of the current steps. There is a strong relation between the oscillations of the background modes and the
static kink configuration: the opposite static kink chains in the different
junctions correspond to opposite ($\pi$-shifted) oscillations of the background modes.

\section{Mathematical Model}
Assuming zero mode along y-axis
we consider a system of 1D Sine-Gordon equations ($xz$-model) \cite{p16}.
For stack of $N$ periodically stacked junctions
with length $L$ the column vector $\varphi=(\varphi_1,...,\varphi_N)^T$ of
the gauge invariant phase differences
satisfies  the standard inductive model \cite{p17}:
\begin{equation}\label{e2}
S(\varphi_{tt}+\alpha \varphi_{t}+ \sin \varphi-\gamma)=\varphi_{xx}, \ \ 0<x<L.
\end{equation}
Here $S$ is the $ N\times N$  matrix
$$S=\left(
\begin{array}{cccccc}
  1 & s & 0 & . & 0 & s \\
  s & 1 & s & 0 & . & 0 \\
  . & . & . & . & . & . \\
  . & . & . & . & . & . \\
  0 & . & 0 & s & 1 & s \\
  s & . & 0 & 0 & s & 1
\end{array}
\right)$$
with $s = - \lambda / (w\sinh (d/\lambda) + 2 \lambda \cosh(d/\lambda))$ being the inductive coupling parameter,
$\lambda$ is the London penetration depth, $w$ is the thickness of insulators, $d$ - the thickness of superconductors.

In \eqref{e2} $\alpha=\sqrt{\hbar/2e j_c C R^2}$ is the damping parameter, where
$R,C,j_c$ are the normal resistance, the capacitance and the critical current per unit length respectively.
The applied bias current $\gamma$  is normalized to the critical current $j_c$.
The space $x$ in \eqref{e2} is normalized to the Josephson penetration depth $\lambda_J=\sqrt{\hbar / 2e\mu_{0}j_ct^{'}}$, the
time $t$ to the inverse of plasma frequency $\omega_{pl}=\sqrt{\hbar C /2ej_c}$, where  $t^{'}$ is the effective magnetic thickness of the barrier,
$\mu_{0}$ is the magnetic constant.
In calculated further current-voltage characteristics (CVCs) the voltage $V$ will be normalized to $\hbar\omega_{pl}/2e$.
The voltage $V$ in the stack is the sum of voltages $v_i$ in the junctions: $V=\sum_{i=1}^{N} v_i$.
When $v_i$ are equal as a result of using  periodic conditions for the stack, we will write simply $v$.

Together with  equation \eqref{e2} we consider radiative boundary conditions \cite{p18},\cite{p19}:

\begin{equation}\label{e3}
\varphi_{x} (x=0,L)=\pm\frac{1}{\widetilde{Z}}(\varphi_{t}(x=0,L)-\langle \varphi_{t}\rangle).
\end{equation}

With $\langle .\rangle$ we denote the instant spatial average, plus sign corresponds to the left edge of the stack ($x=0$), minus sign - to the right edge ($x=L$). $\widetilde{Z}$  is the normalized impedance.
For $\widetilde{Z}=\infty$  boundary conditions \eqref{e3} become simple Neumann conditions:
\begin{equation}\label{e4}
\varphi_{x} (x=0,L)= 0.
\end{equation}
Because $\widetilde{Z}$ is huge in reality \cite{p13} and also it is not easy to implement \eqref{e3}, most authors prefer to use Neumann conditions for their simulations.
In our simulations we use the boundary conditions \eqref{e3}.

\section{Cavity mode frequencies and the coupling parameter}

For rectangular mesas described by the $xz$-model, the frequencies for the cavity modes $m,m=1,2,3...$ (in normalized units) are given by \cite{p20}:
\begin{equation}\label{e5}
{\omega^m_{cav}}=\frac{m C_{max}}{2L}
\end{equation}
Here $C_{max}$ is the maximum plasma velocity. $C_{max}=1/\sqrt{\lambda_{min}(S)}= 1/\sqrt{(1+2s)}$, where
$\lambda_{min}(S)$ is the minimal eigenvalue of the cyclic tridiagonal matrix $S$ and does not depend on $N$.
For the usual (not cyclic) tridiagonal matrix $\lambda_{min}= 1/\sqrt{(1+2s\cos(\pi/(N+1))}$ \cite{p13}.

In our simulations we take $s$ close to $s=-0.5$, which corresponds to strong inductive coupling (as for BSCCO crystals).
It is important to note again the difference between using models with cyclic and not cyclic matrix.
As an example for $s=-0.49985, N=12$, $C_{max}$ for the cyclic and not cyclic matrix are respectively $57.7$ and $5.83$,
and the condition numbers of the  matrices are respectively $6665$ and $67.1$. For $s=-0.499995$, which is  usually used in simulations \cite{p21},
$C_{max}$ changes to $316$ and $5.87$  respectively and the condition numbers to $2.10^{5}$ and $67.8$.
The above difference in wave velocities  nicely explains the  different results in \cite{p11},\cite{p13} and the  achieved higher modes in \cite{p13} - up to 12.
Moreover, it is difficult to determine  parameter $s$ rigorously \cite{p22}.
For all the above reasons (the use of periodic boundary conditions, numerical accuracy, not strictly determined $s$ in reality),
we  take $s$ phenomenologically. The results that follow are achieved for $s=-0.49985$.

\section {Results}

We solve problem \eqref{e2},\eqref{e3} by using a leap-frog difference scheme. The nonstandard  boundary  conditions \eqref{e3} are approximated by the method of balance \cite{p23},
which provides energy conservation of the numerical scheme.
Simulations are made for $N=12, L=30, s=-0.49985, \alpha=0.05,0.07,0.13, \widetilde{Z}=6.10^5$. The bias current $\gamma$ is taken with  noise of amplitude $10^{-8}$.

\begin{figure}[ht]
\includegraphics[width=0.36\textwidth]{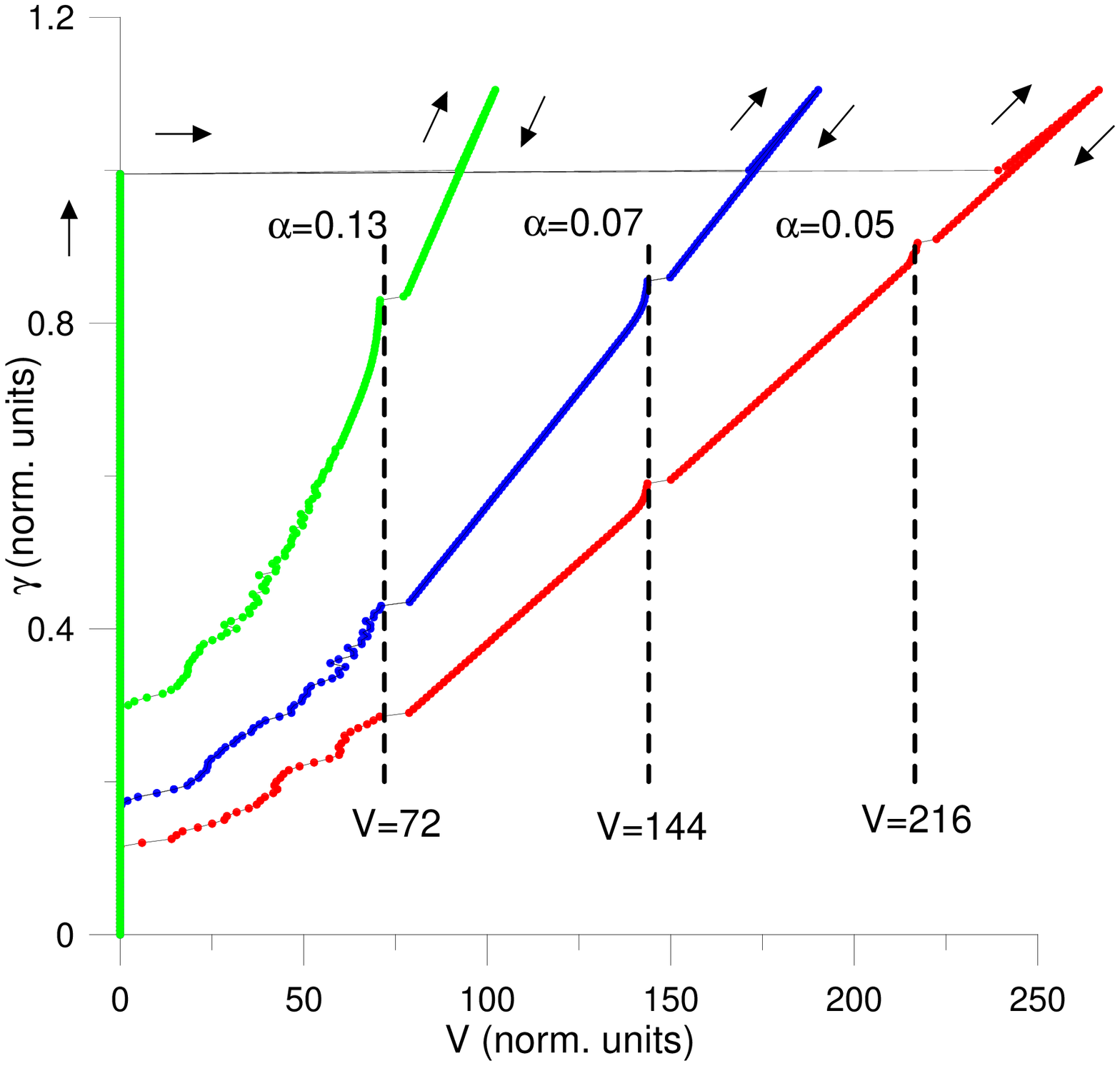}
\caption{\label{fig: 1} Simulated CVC's for different dissipation parameters.}
\end{figure}
\begin{figure}[ht]
\includegraphics[width=0.36\textwidth]{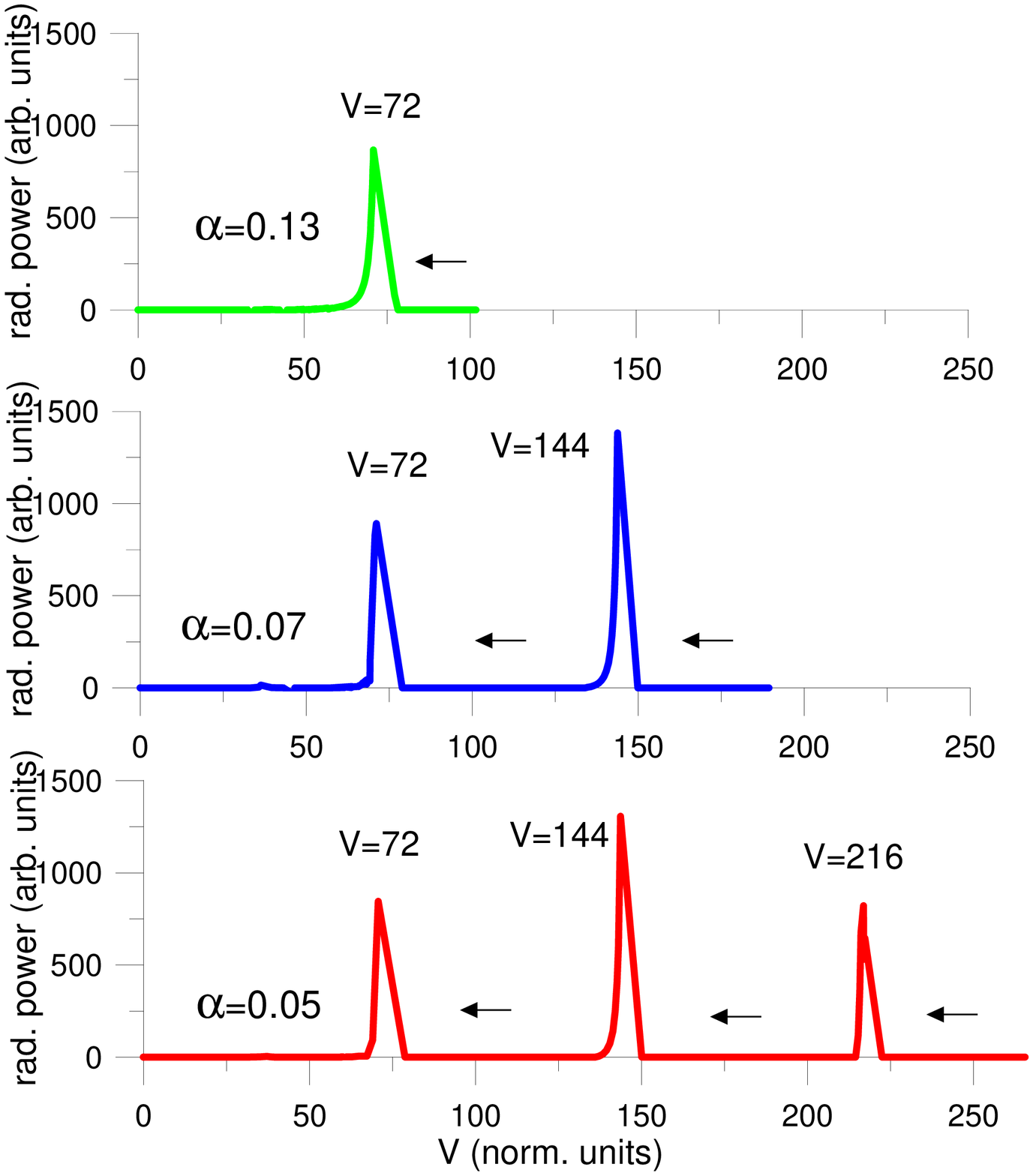}
\caption{\label{fig: 2} Emission power corresponding to CVC's  calculated by the Poynting vector at the edge $x=0$.}
\end{figure}

We don't use any special initial conditions. We use zero initial conditions at the start of our simulations.
A typical simulation of the outermost branch of the CVC works as follows. Starting with zero initial conditions at $\gamma=0$ we increase $\gamma$ by steps of $0.005$ until (at $\gamma=1.0$)  all the junctions switch to resistive state. We continue increasing $\gamma$ up to $1.1$  and then start to decrease $\gamma$.
A hysteresis region, typical for IJJ, is observed, i.e. at the decreasing part of CVC, the voltage $V$  remains nonzero for $\gamma<1$ until all the junctions switch to superconductive state ($V=0$)
at some bias current.

CVCs for $\alpha=0.05,0.07,0.13$ are shown in Fig.1. Different $\alpha$ model different mesa temperatures \cite{p13}.
CVCs show a nice regular pattern. Depending on the possible achievable voltage $V$, jumps are observed at $V\approx72, V\approx144=2\times72,  V\approx216=3\times72$.
Voltage $V\approx72$ in the stack corresponds to voltage $v\approx 6$ per junction.
Radiation power achieved by calculating time average of the Poynting vector \cite{p24} at the edge $x=0$ of the stack is shown in Fig.2.
Each voltage jump corresponds to a peak in the radiation power.
These voltage jumps are different from the jumps to another voltage branch such as  those in the experimental results in \cite{p1}, where different groups of junctions are involved in resonance.
In our case we  remain on the same (outermost) branch.

Direct analysis  of the calculated $\varphi$-dynamics shows that the peaks at $V\approx72,144,216$ correspond to an excitation of the cavity modes $1,2,3$ respectively \cite{p20}.
The phase  in a particular junction is a sum of three terms: the linear term $vt$, a static kink term (time independent) and an oscillating term \cite{p11}.
A clear two-dimensional standing wave pattern is seen as in \cite{p11},\cite{p13}, where
the oscillating term of the phases $\varphi$ is approximately uniform in z-direction and  satisfies the linear wave equation:
\begin{equation}\label{e6}
\varphi_{tt}={{(C_{max})}}^2\varphi_{xx}
\end{equation}
with Neumann boundary conditions, except at the bottom of the current steps.
As our numerical simulations show there exist a small nonuniform oscillating part with the same frequency, which is clearly seen at the bottom of the current steps. Unlike the cavity mode part this part has a nonlinear origin. We use the terminology from \cite{p10} and call this oscillating mode a background mode. We will return to describe the background modes later.

It is  worth to note that the cavity mode 1 excitation is more stable for larger dissipation parameter $\alpha$.
This is clearly seen from the loosing of smoothness  of CVC around the resonance voltage for $\alpha=0.05,0.07$ (Fig. 1) and also from the direct observation
of the phase $\varphi$ which shows a very complicated picture, hard to analyze.
This  is in a good agreement with the experimental results in \cite{p4},\cite{p5} where the highest power for cavity mode 1
is achieved for relatively higher temperatures.

\subsection{\label{sk}Static kink configurations}

The static part of $\varphi$ consists of some configuration of chains of kinks ($k$) and anti-kinks ($a$) in every junction as in \cite{p11},\cite{p12}.
A crucial observation is the alternative character of the static kink configuration, i.e. alternating of opposite kinks (kinks and anti-kinks)  in x and z-direction. As numerous simulations show the  number of kinks and antikinks in the stack is usually one and the same; if it is not, the difference is  at most 2. For mode $m$ the chains consist of $m$ kinks or antikinks, for example $\underbrace{akkaak...ak}_m$.
A kink configuration in the stack is a sequence of chains starting with the first junction. For example in the case of excitation of third cavity mode the sequence $(aka, kak, aka,...)$
corresponds to a configuration with chain $aka$ in the first junction, $kak$ in the second and so on.
\begin{figure*}
\centering
\begin{minipage}[b]{.45\textwidth}
  \centering
  \includegraphics[width=0.70\textwidth]{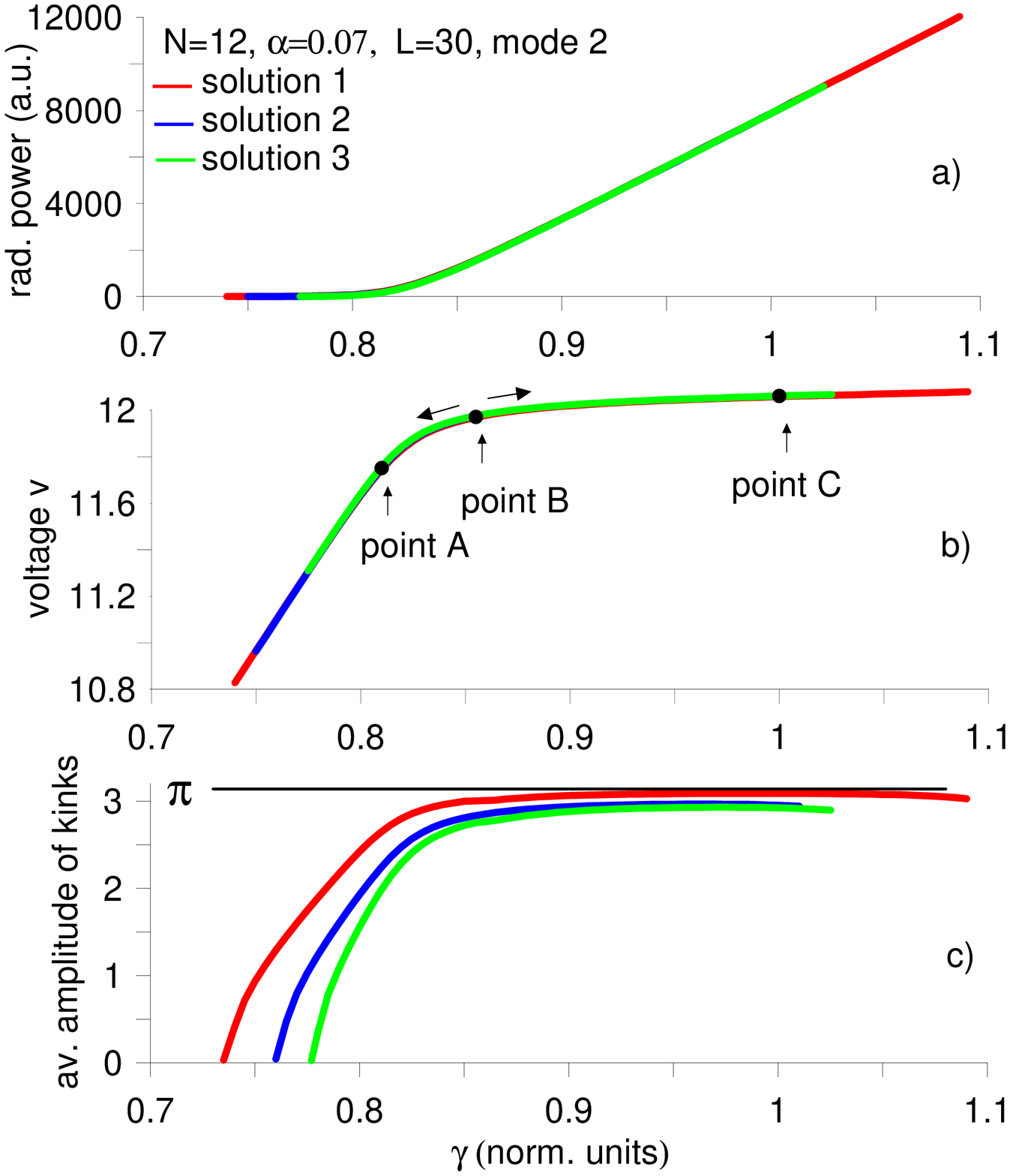}
  \caption{\label{fig: 3} Kink amplitudes, current-voltage branches
  and emission powers for different kink configurations corresponding to excitation
  of cavity  mode 2.}
\end{minipage}\qquad
\begin{minipage}[b]{.45\textwidth}
  \centering
  \includegraphics[width=0.74\textwidth]{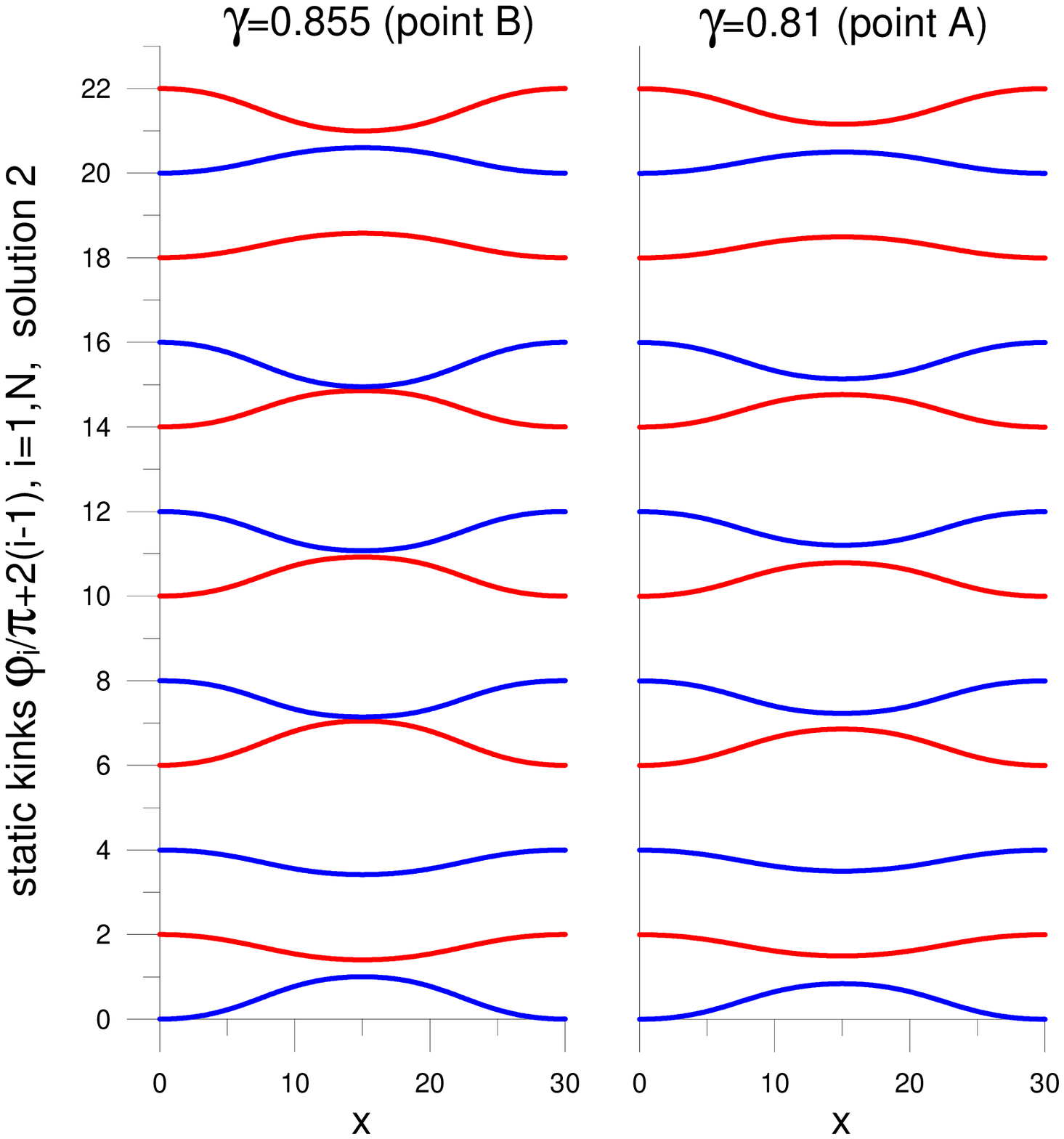}
  \caption{\label{fig: 4}Static kink configuration
   at points $A$ and $B$ \\ for solution 2.\\ }
\end{minipage}
\end{figure*}

\begin{figure}[ht]
\includegraphics[width=0.36\textwidth]{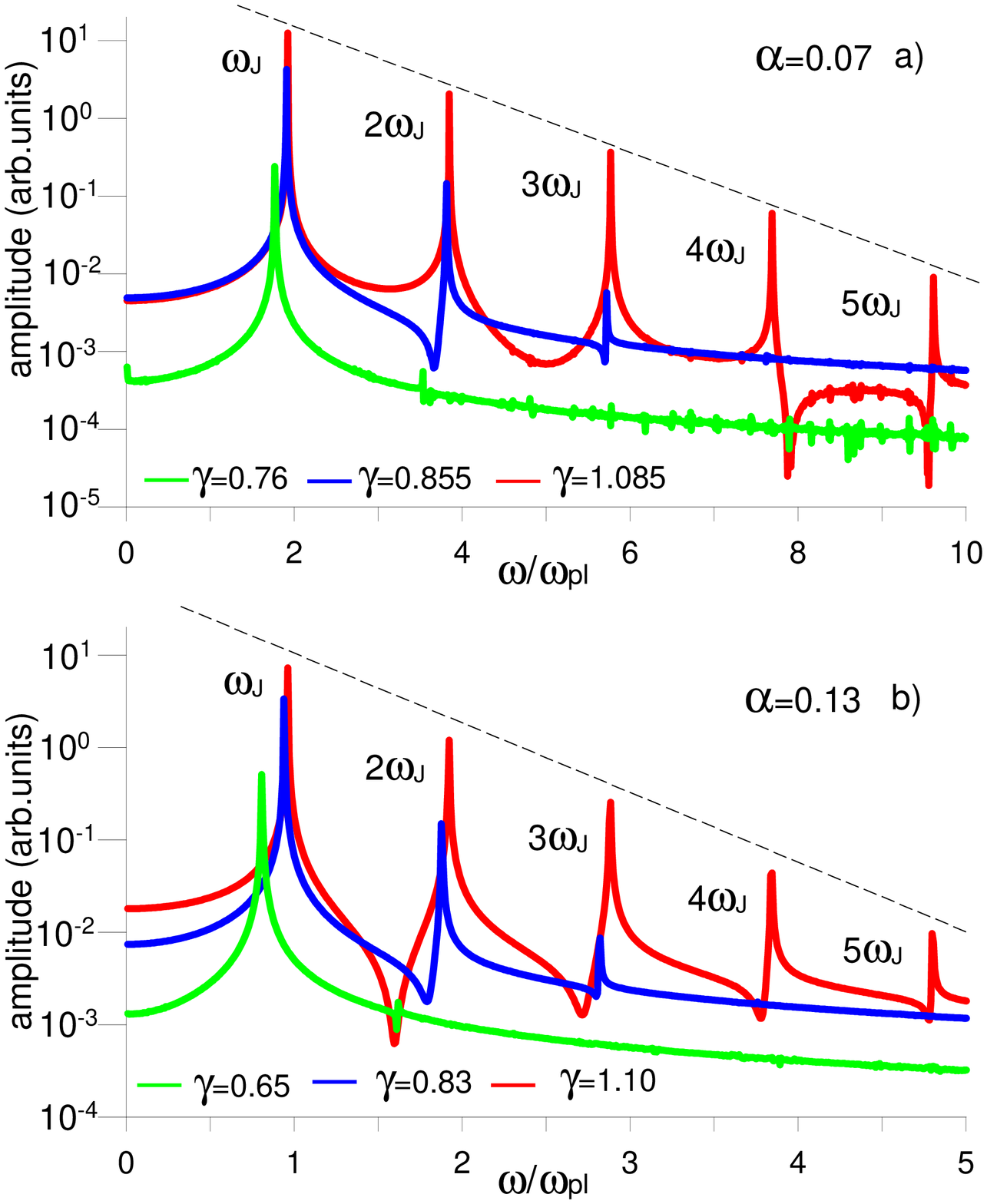}
\caption{\label{fig: 5}
    FFT of the electric field $\varphi_t$ at the edge $x=0$ of the stack for different bias points
  on the current steps of cavity mode 2 (a) and cavity mode 1 (b).}
\end{figure}

We analyze now in  detail   some kink configurations observed when  cavity mode $2$  is excited. They are achieved at $N=12, L=30, s=-0.49985, \alpha=0.07$.
The picture for cavity mode $1$, achieved at $N=12, L=30, s=-0.49985, \alpha=0.13$ and for mode 2 at different $\alpha$ is similar.
Regardless the static kink configuration for numerous repeated simulations, cavity mode $2$ is excited always when $\gamma=0.855$ at the decreasing part of CVC.
The results for the radiation power, the voltages and the kink amplitudes, corresponding to three different kink configurations (three solutions), are shown in Fig.3.
All the curves in Fig.3 are obtained by reversing $\gamma$ at point $B$ ($\gamma=0.855$), i.e. increasing $\gamma$ until the corresponding solution exists.
Figure 3b actually shows the current steps in the CVC corresponding to the excitation of cavity mode 2 with different kink configurations.
The heights of the current steps differ for different configurations.
Solution 1 corresponds to the kink configuration $(ka,ak,ka,ak,ka,ak,ka,ak,ka,ak,ka,ak)$,
solution 2 - to $(ka,ak,ak,ka,ak,ka,ak,ka,ak,ka,ka,ak)$,
solution 3 - to $(ka,ak,ak,ka,ka,ak,ak,ka,ak,ka,ka,ak)$.
The total number of kinks and antikinks for the above kink configurations is one and the same.
From the top to the bottom of the current steps  the kink amplitudes decrease from almost $\pi$ to zero.
Solution 1 exists at largest range in $\gamma$ and can achieve maximal radiation power (Fig.3a).
Although for every coexisting point the voltages and the power are approximately the same for solutions $1,2,3$, the kink amplitudes are not.
The average in the stack kink amplitudes are shown in Fig.3c. The average kink amplitude is greatest for Solution 1 for every $\gamma$.
The maximal average kink amplitude is close ($\approx 3.10$), but less then $\pi$. Decreasing $\gamma$ from point $B$ the kink amplitudes strongly decrease together
with decreasing the voltage and the radiation power.
Kink configuration for solution 2 at point $A$ ($\gamma=0.81$) and point $B$ ($\gamma=0.855$) in Fig.4 confirm the last conclusion.
It is worth to note that although the average kink amplitude is always less then $\pi$, it could be even greater then $\pi$ for particular junction and depending on the kink configuration.
For example the kink amplitude in junction $9$ for  solution 3 at point $C$ ($\gamma=1.0$) is $3.2631$.

\begin{figure}[ht]
\includegraphics[width=0.33\textwidth]{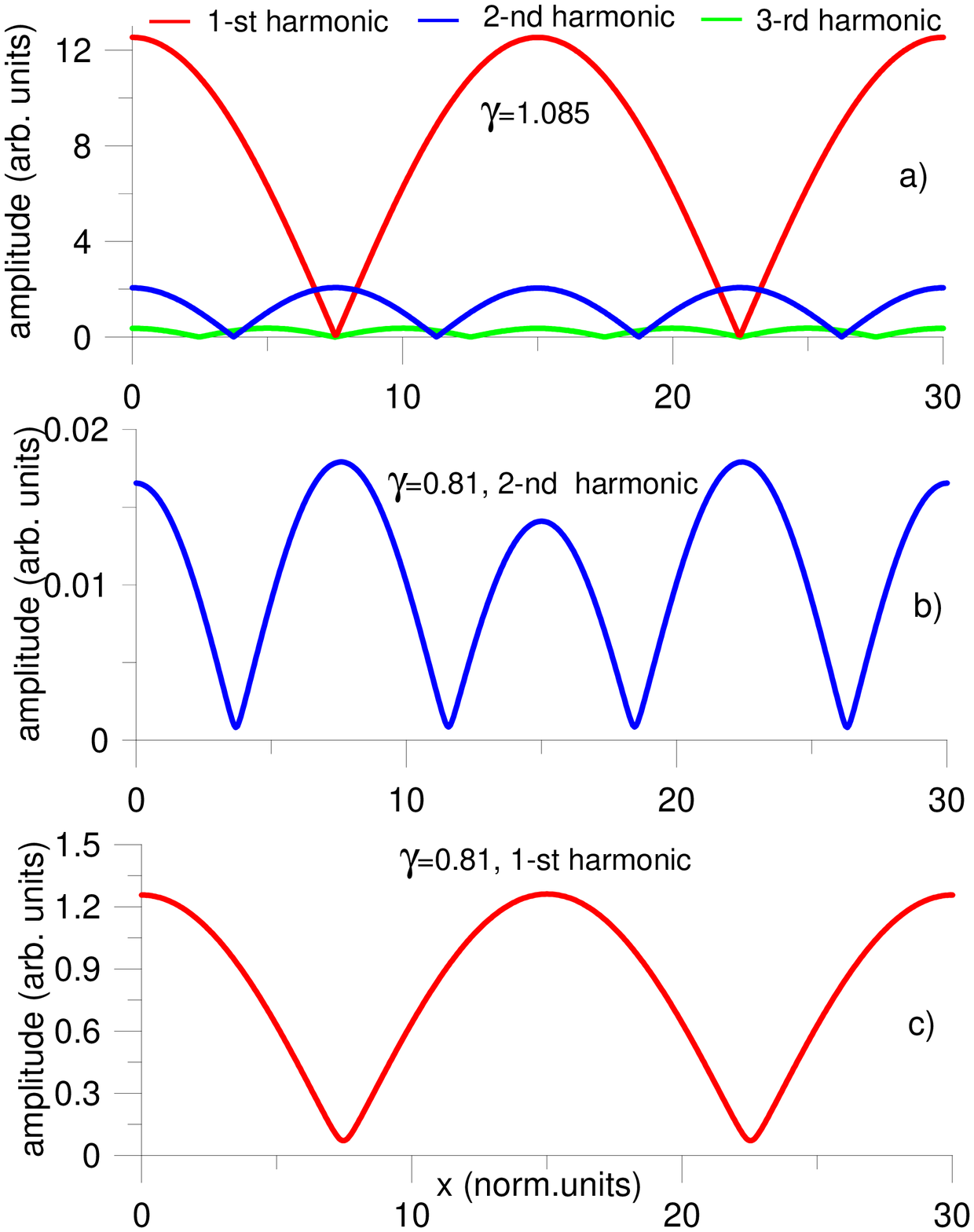}
\caption{\label{fig: 6}  Amplitude maps of the electric field at different bias currents:
  At the top of the current step (a) the amplitudes for the first three harmonics excellently fit to the cavity amplitudes.
  At the bottom of the current step (b,c) the amplitudes deviates from those corresponding to the cavity modes.\\}
\includegraphics[width=0.33\textwidth]{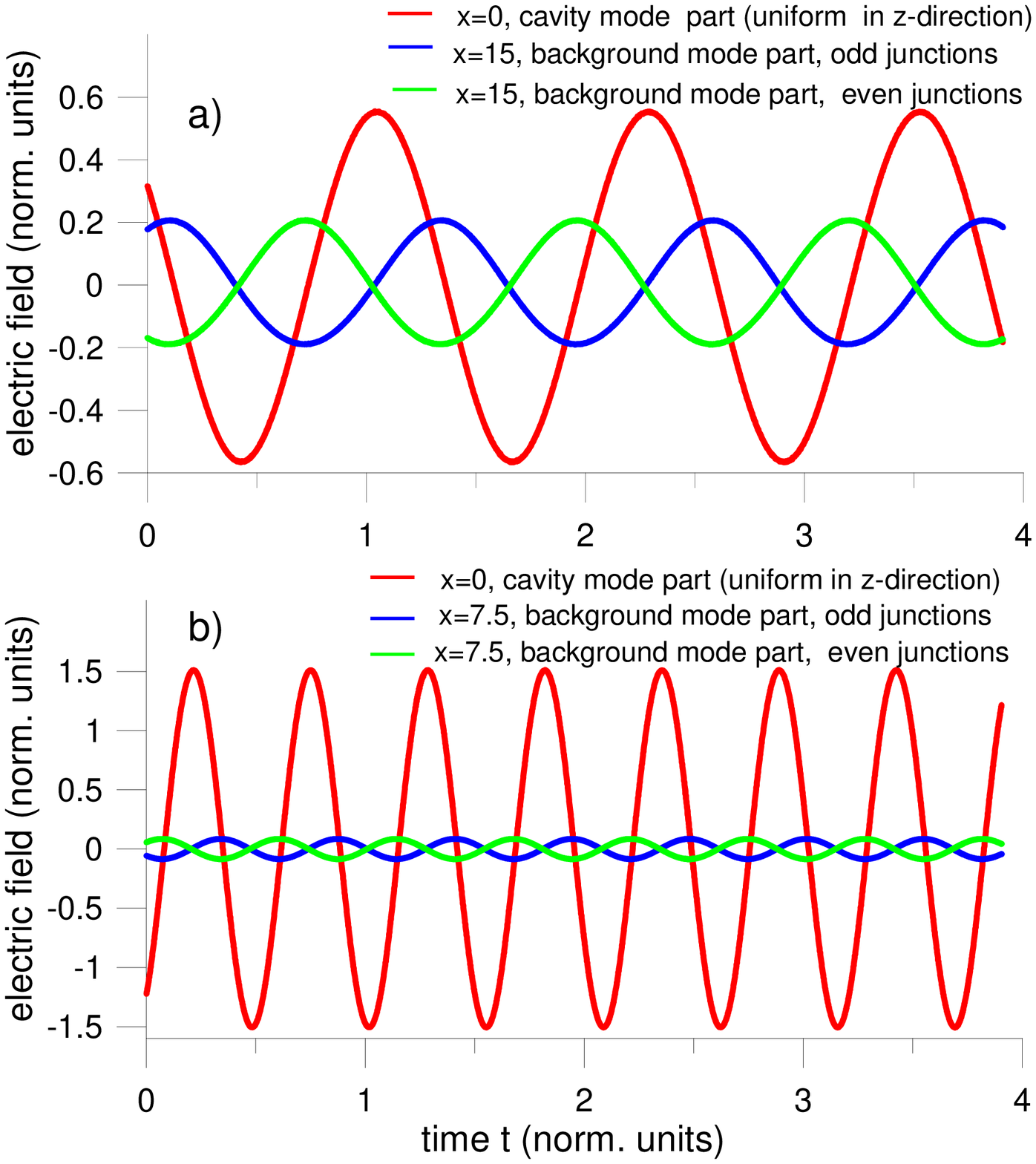}
\caption{\label{fig: 7} Oscillations of  the cavity mode part and the background mode part of the electric field
  for the excited cavity mode 1 (a) and cavity mode 2 (b). The shift between  the oscillations of the background and the cavity mode is  close to $\pi/2$.
  The oscillations of the background mode in the adjacent junctions are anti-phase ($\pi$ - shifted).}
\end{figure}

\begin{figure*}
\centering
\begin{minipage}[b]{.45\textwidth}
  \includegraphics[width=0.9\textwidth]{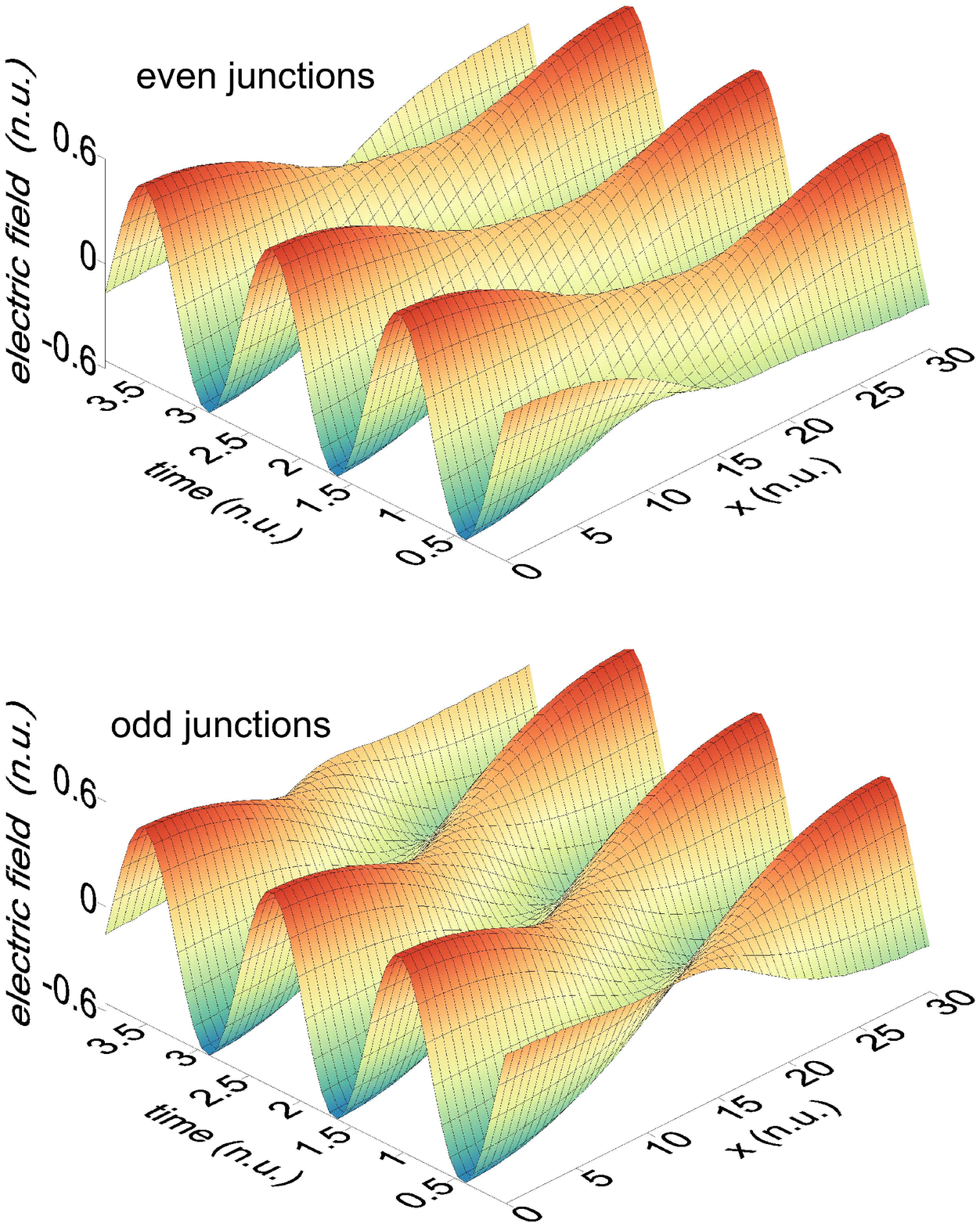}
  \caption{\label{fig: 8} Time evolution of the electric field $\varphi_t$ at the bottom of the current step corresponding
   to an excitation of cavity mode 1. Parameters are $\alpha=0.13$, $\gamma=0.65$. The existence of the background mode is visible. }
\end{minipage}\qquad
\begin{minipage}[b]{.45\textwidth}
  \includegraphics[width=0.9\textwidth]{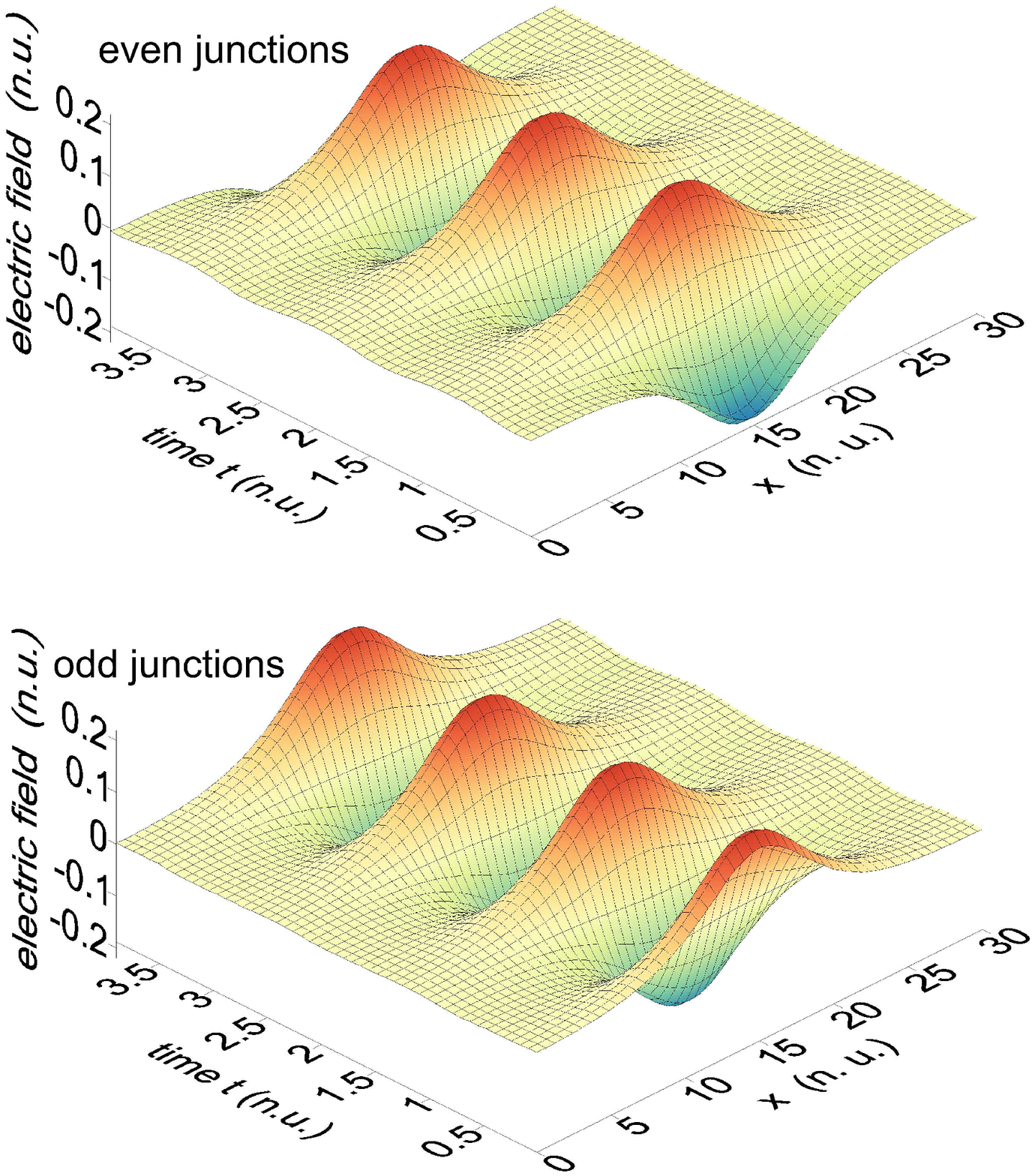}
  \caption{\label{fig: 9} Time evolution of the background mode part of the electric field (the electric field $\varphi_t$ with subtracted cavity mode part). Parameters are $\alpha=0.13$, $\gamma=0.65$.
   Oscillations in adjacent junctions are anti-phase ($\pi$-shifted). }
\end{minipage}
\end{figure*}
\begin{figure*}
\centering
\begin{minipage}[b]{.45\textwidth}
  \includegraphics[width=0.9\textwidth]{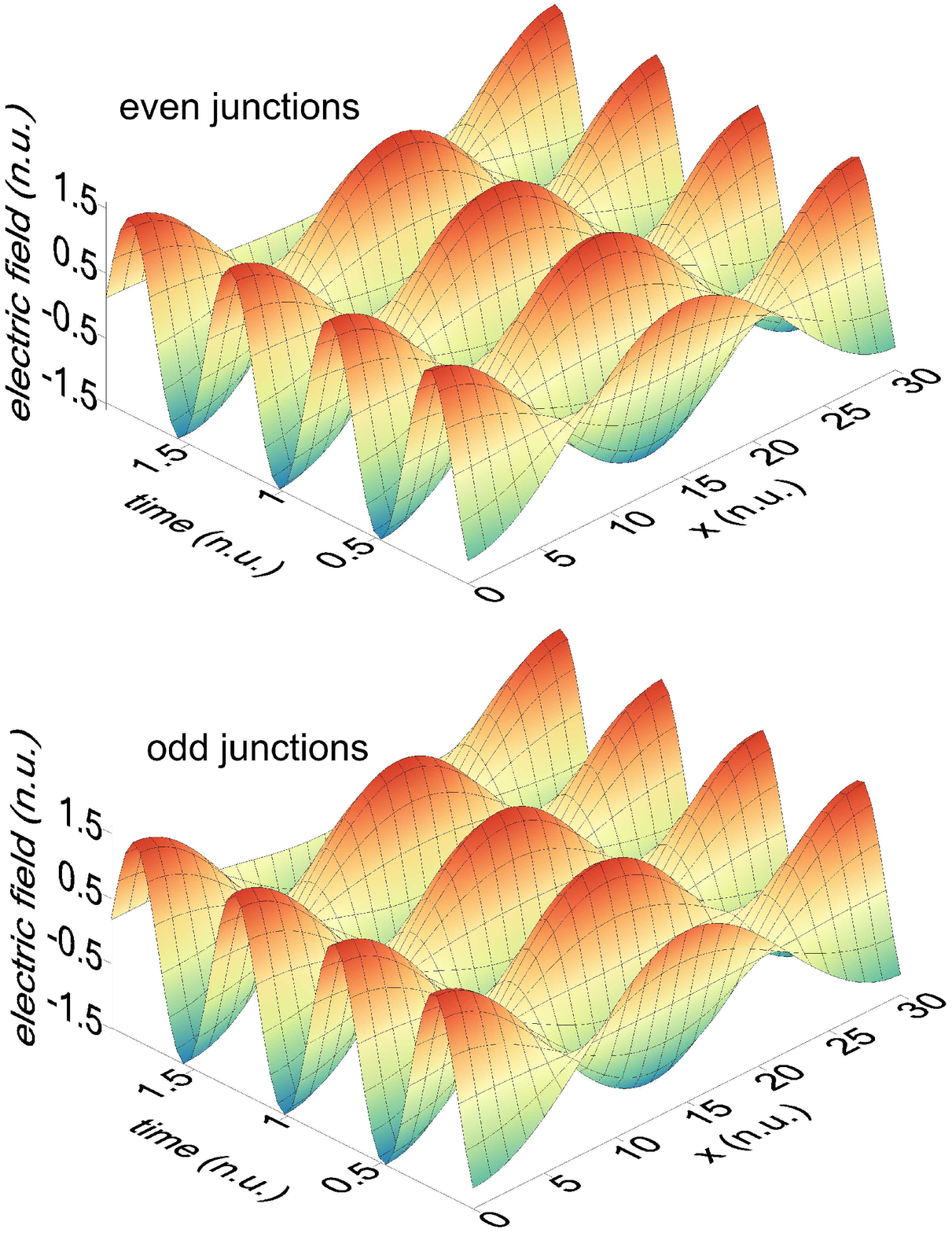}
  \caption{\label{fig: 10} Time evolution of the electric field $\varphi_t$ at the bottom of the current step corresponding
   to an  excitation of cavity mode 2 . Parameters are $\alpha=0.07$, $\gamma=0.81$. The existence of the background mode is not clearly visible because its amplitude is relatively small.}
\end{minipage}\qquad
\begin{minipage}[b]{.45\textwidth}
  \includegraphics[width=0.9\textwidth]{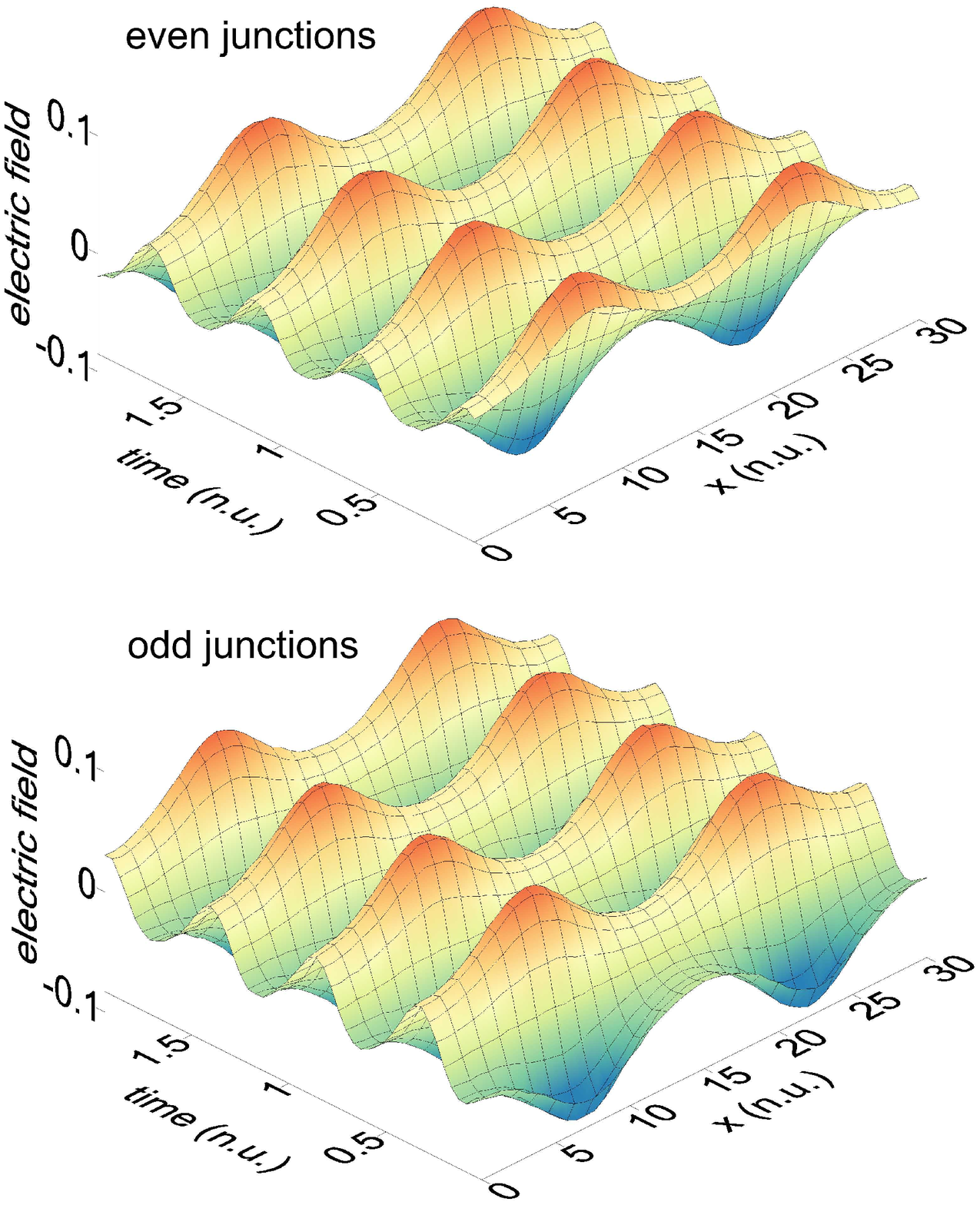}
  \caption{\label{fig: 11}  Time evolution of the background mode part of the electric field (the electric field $\varphi_t$ with subtracted cavity mode part and its second harmonic).
   Parameters are $\alpha=0.07$, $\gamma=0.81$. Oscillations in adjacent junctions are anti-phase ($\pi$-shifted).}
\end{minipage}
\end{figure*}

\subsection{\label{fft}  Fast Fourier Transform analysis and observation of oscillating background modes }

To analyze the radiation frequency we have used Fast Fourier Transform (FFT) of the electric field $\varphi_t$ with resolution  $2^{-7}\omega_{pl}$.
Calculated by FFT radiation frequency $\omega$  excellently fulfills the ac Josephson relation, which in normalized units is $\omega=\omega_J=v/2\pi$.
The FFT results for the oscillation of the electric field  $\varphi_t$ at the edge $x=0$ of the stack  are shown in Fig.5a (cavity mode 2) and Fig.5b (cavity mode 1).
The oscillation amplitudes decrease together with the decreasing bias current $\gamma$ (decreasing radiation power).
Integer higher harmonics up to the fifth order are also visible on the top of the current steps but their number decrease with  $\gamma$
and completely disappear at the bottom of the current steps. At the top of the current steps the amplitudes decrease exponentially as it is seen from dashed lines in Fig.5.
Integer higher harmonics are obtained  in many experiments,
for example in: \cite{p6},\cite{p8},\cite{p9},\cite{p25},
where an exponential rate of decreasing of harmonics amplitudes is also observed.

To check the resonance condition \eqref{e1} we calculate $\omega_J$  from calculated voltage at the top of the current step of cavity mode 2 and
${\omega^2_{cav}}$ from formula \eqref{e5}. They differ by less then $10^{-5}$: $\omega_J={\omega^2_{cav}}= 1.9245$.
Although the voltage $v$ is in a pretty large range for kink solutions ($v\in (10.8,12)$ in Fig. 3b),
the frequency tunability for powers greater then $10\%$ and $5\%$ of the maximal power is only about $1\%$ and $2\%$ respectively.
The fact that the frequency at the maximal achieved power excellently fits with the cavity frequency agrees with the experimental results \cite{p2}-\cite{p4}.
On the other hand the frequency tunability around the cavity frequency is much larger in the experiments then in our simulation.
In our opinion this is dew to the fact that the considered model doesn't treat the  temperature inhomogeneities which exist in reality.

In order to understand whether the oscillating part consists only of  cavity mode oscillations,
we calculate the amplitude maps of the electric field, i.e. we calculate the amplitude for every harmonic frequency by using  FFT  for every point $x\in [0,L]$.
It is correct because although the amplitudes slightly depend on the output bins, the frequencies are the same for every $x$.
Figure 6a shows  the amplitudes of the first three frequency harmonics corresponding to the top of the current step of the excited cavity mode 2.
In our case the second and the third harmonics
are the cavity modes $4$ and  $6$ respectively. The amplitudes excellently fit  with the properly scaled $|cos|$-function: $|cos(2\pi x/L)|,|cos(4\pi x/L)|,|cos(6\pi x/L)|$.
On the other hand at the bottom of the current step (Fig.6b,c) the amplitudes  deviate from those corresponding to the cavity modes, particularly they
are far from zero at the nodes of the cavity modes. This result confirms the existence of oscillating background modes.
Although it is hard to detect background mode oscillations at the top of the current steps, we suppose that
they support the cavity mode oscillations at the entire current step. The background modes have a nonlinear origin.
Direct analysis shows that they are strongly related with the static kink configurations: opposite  static kink chains in different
junctions correspond to opposite ($\pi$-shifted) oscillations of the background modes.
In our case the opposite static kink chains for the cavity mode 1 are $a$ and $k$, for the cavity mode 2 they are $ak$ and $ka$.

The oscillations of the background modes in the adjacent junctions (odd and even) for excited cavity mode 1 at $\gamma=0.65$, $\alpha=0.13$ are
shown in Fig.7a. The static kink configuration is $(k,a,k,a,k,a,k,a,k,a,k,a)$. The background oscillations in the odd and the even junctions are anti-phase
($\pi$-shifted  to one another). The shift with respect to the cavity mode oscillation slightly deviates from $\pi/2$.
The time evolution of the electric field  and the electric field with subtracted cavity mode part are shown in Fig.8 and Fig.9 respectively.
As it is seen from these figures, the amplitudes of the background modes are not uniform, but they are higher at the centers of the kinks.

The oscillations of the background modes in the adjacent junctions (odd and even) for excited cavity mode 2 at $\gamma=0.81$, $\alpha=0.07$ are
shown in Fig.7b. The static kink configuration is $(ka,ak,ka,ak,ka,ak,ka,ak,ka,ak,ka,ak)$ (solution 1).
Here the background amplitudes are relatively smaller because we are not at the very bottom of the current step.
The picture is similar, the  anti-phase background oscillations are clearly seen in Fig. 7b.
The shift with respect to the cavity mode oscillation is very close to $\pi/2$.
The time evolution of the electric field and the electric field with subtracted first and second cavity harmonics are
shown in Fig.10 and Fig.11 respectively. Background mode oscillations are again nonuniform, but higher at the centers of the kinks.
Let us note that if we consider  kink configuration $(ka,ak,ak,ka,ak,ka,ak,ka,ak,ka,ka,ak)$ (solution 2),
the background oscillations are not anti-phase in the adjacent junctions (odd and even), but they are anti-phase in the following groups of junctions: \{1,4,6,8,10,11\}
and \{2,3,5,7,9,12\}.

\section {Conclusions}
By means of numerical  simulations we show that the coupling to the resonance modes is obtained by a static kink configuration in z-direction.
The kink amplitudes may differ from $\pi$ depending on the kink configuration and the point on the current step.
The kink configurations have alternative character with approximately equal number of kinks and anti-kinks.
Calculated amplitude maps for the radiation frequencies and direct phase observation show
the existence of nonlinear background modes together with the cavity modes, thus contributing to the improvement of the mechanism for powerful THz radiation in IJJs at zero magnetic field.
A relation between the oscillations of the  background modes and the static kink configuration is established:
opposite static kink chains in different
junctions correspond to opposite ($\pi$-shifted) oscillations of the background modes.

\begin {center}
  \textbf{Acknowledgments}
\end {center}
We thank Prof. D.Sc. Shukrinov and Dr. Rahmonov
for valuable discussions and important remarks.
This work is supported by the National Science Fund of BMSE under grant I-02/9/2014
and by the Sofia University Science Fund under grant N75/2015.
We thank for the opportunity to use the computational resources of IICT-BAS.

\end{document}